\documentclass[showpacs,amsmath,amssymb]{revtex4}

\usepackage{graphics}
\usepackage[dvips]{graphicx}

\def\sqr#1#2{{\vcenter{\hrule height.#2pt\hbox{\vrule width.#2pt height#1pt
\kern#1pt \vrule width.#2pt}\hrule height.#2pt}}}
\def\square{\mathchoice\sqr64\sqr64\sqr{4.2}3\sqr{3.0}3}

\begin{document}

\title{Cosmological simulations: the role of scalar fields}


\author{M.A. Rodr\'{\i}guez-Meza}

\address{Depto. de F\'{\i}sica, Instituto Nacional de Investigaciones Nucleares,
  Col. Escand\'on, Apdo. Postal 18-1027, 11801 M\'{e}xico D.F. \\
 marioalberto.rodriguez@inin.gob.mx; http://astro.inin.mx/mar
  }

\date{\today}


\begin{abstract}
We present numerical $N$-body simulation studies of large-scale structure formation.
The main purpose of these studies is to analyze the several models of dark matter
and the role they played in the process of large-scale structure formation. We analyze
the standard and more successful case, i.e., the cold dark matter with cosmological
constant ($\Lambda$CDM). We compare the results of this model with the corresponding results
of other alternative models, in particular, the models that can be built from the Newtonian
limit of alternative theories of gravity like scalar-tensor theories. An specific model
is the one that considers that the scalar field is non-minimally coupled to the Ricci
scalar in the Einstein-Hilbert Lagrangian that gives, in the Newtonian limit
an effective gravitational force that is given by two contributions: the standard
Newtonian potential plus a Yukawa potential that comes from a massive scalar field.
Comparisons of the models are done by analyzing 
the snapshots of the $N$-body system
at z=0 for several values of the SF parameters. 
\end{abstract}

\maketitle


\section{Introduction}
In this work we present some results about the role scalar fields (SF) play 
in cosmological simulations, in particular
on the process of large scale structure formation.
The main goal of this work is to study the large scale structure formation 
where  the usual approach is that the evolution of 
the initial primordial fluctuation energy density fields evolve following 
Newtonian mechanics in an expanding background\cite{Peebles1980}.
The force between particles are the standard Newtonian gravitational force.
We will see that we can introduce SF by adding a term in this force.
This force will be of Yukawa type with two parameters 
($\alpha$, $\lambda$)\cite{mar2004}.
We have been studying, in the past years, the effects of 
this kind of force on some astrophysical 
phenomena\cite{mar2004,Rodriguez2001,mar2005,jorge2007}.
The Yukawa force comes as a Newtonian limit of a scalar-tensor theory (STT) with the SF
non-minimally coupled to gravitation\cite{mar2004}.
It is our purpose to find the role these scalar fields play on the large scale structure 
formation processes.
We start by discussing the standard $\Lambda$CDM model and the general approach in $N$-body
simulations (See Bertschinger\cite{Bertschinger1998} for details). 
Then, we present the modifications
we need to do to consider the effects of a static SF and we show the
results of this theory for the cosmological concordance model of a $\Lambda$CDM 
universe\cite{Heitmann2005}. To perform the simulations we have modified a standard serial 
treecode the author has developed \cite{Gabbasov2006} and the Gadget 1 \cite{Springel2001} 
(see also \url{http://www.astro.inin.mx/mar})
in order to take into account the contribution of the Yukawa potential.

\section{Evolution equations for a $\Lambda$CMD universe}
\subsection{General Scalar-tensor theory and its Newtonian limit}
Let us consider the Einstein field equations of a typical STT\cite{Faraoni2004}
\begin{eqnarray}
R_{\mu\nu} - \frac{1}{2} g_{\mu\nu} R &=& \frac{1}{\phi}
\left[ 8 \pi T_{\mu\nu} + \frac{1}{2} V g_{\mu\nu}
+ \frac{\omega}{\phi} \partial_\mu \phi \partial_\nu \phi
\right. \nonumber \\
&& \left. -\frac{1}{2} \frac{\omega}{\phi}(\partial \phi)^2 g_{\mu\nu}
+ \phi_{;\mu\nu} - g_{\mu\nu} \, \square \phi \frac{\mbox{}}{\mbox{}}
\right] \; , 
\end{eqnarray}
for the metric $g_{\mu\nu}$ and for the massive SF $\phi$ we have
\begin{equation}\label{EqSTTPhi}
\square \phi  = \frac{1}{3+2\omega} \left[
	8\pi T -\omega' (\partial \phi)^2 +\phi V' - 2V \right] \, ,
\end{equation}
where $()' \equiv \frac{\partial }{\partial \phi}$. Here $T_{\mu\nu}$ is the energy-momentum
tensor with trace $T$, $\omega(\phi)$ and $V(\phi)$ are in general arbitrary functions that 
gobern kinetic and potential contribution of the SF. The potential contribution, $V(\phi)$,
provides mass to the SF, denoted here by $m_{SF}$.

The study of large-scale formation in the universe is greatly simplified by the fact that a
limiting approximation of general relativity, Newtonian mechanics, applies in a region
small compared to the Hubble length $cH^{-1}$ ($cH_0^{-1}\approx 3000 h^{-1}$ Mpc, where 
$c$ is the speed of light, $H_0=100 h$ km/s/Mpc, 
is Hubble's constant and $h\approx (0.5-1)$), and large compared to the Schwarzschild
radii of any collapsed objects. The rest of the universe affect the region only through a tidal field.
The length scale $cH_0^{-1}$ is of the order of the largest scales currently accessible in 
cosmological observations and $H_0^{-1} \approx 10^{10}h^{-1}$ yr 
characterizes the evolutionary time scale of the universe.

Therefore, in the present study, we need to consider the influence
of SF in the limit of a static STT, and then we need to describe
the theory in its Newtonian approximation, that is, where gravity 
and the SF are weak (and time independent) and velocities of dark matter
particles are
non-relativistic.  We expect to have small deviations of
the SF around the background field, defined here as
$\langle \phi \rangle$ and can be understood as the scalar field beyond all matter.
If one defines the perturbations $\phi = \langle \phi \rangle + \bar{\phi}$ and
$g_{\mu\nu} = \eta_{\mu\nu} + h_{\mu\nu}$,
where $\eta_{\mu\nu}$ is the Minkowski metric, the Newtonian approximation
gives \cite{mar2004}
\begin{eqnarray}
R_{00} = \frac{1}{2} \nabla^2 h_{00} &=& \frac{G_N}{1+\alpha} 4\pi \rho
- \frac{1}{2} \nabla^2 \bar{\phi}  \; ,
\label{pares_eq_h00}\\
  \nabla^2 \bar{\phi} - m_{SF}^2 \bar{\phi} &=& - 8\pi \alpha\rho \; ,
\label{pares_eq_phibar}
\end{eqnarray}
we have set $\langle\phi\rangle=(1+\alpha)/G_N$ 
and $\alpha \equiv 1 / (3 + 2\omega)$.  In the above expansion we have set
the cosmological constant term equal to zero, since on galactic
scales its influence should be negligible.  We only consider the
influence of dark matter due to the boson field of mass $m_{SF}$ governed by
Eq.\ (\ref{pares_eq_phibar}), that is the modified Helmholtz equation.
However, at cosmological scales we do take into account the cosmological
constant contribution, see below.
Equations (\ref{pares_eq_h00}) and (\ref{pares_eq_phibar}) represent
the Newtonian limit of the STT with arbitrary potential $V(\phi)$ and function
$\omega(\phi)$ that where Taylor expanded around $\langle\phi\rangle$.
The resulting equations are then distinguished by the constants
$G_N$, $\alpha$, and $\lambda=h_P/m_{SF}c$. Here $h_P$ is Planck's constant.

The next step is to find solutions for this new Newtonian potential given 
a density profile, that is, to find the so--called potential--density pairs. 
General solutions to Eqs. (\ref{pares_eq_h00}) and (\ref{pares_eq_phibar}) 
can be found in terms of the corresponding Green functions,
and the new Newtonian potential is\cite{mar2004,mar2005}
\begin{eqnarray}
\Phi_N  \equiv \frac{1}{2} h_{00}
&=& - \frac{G_N}{1+\alpha} \int d{\bf r}_s
\frac{\rho({\bf r}_s)}{|{\bf r}-{\bf r}_s|} \nonumber \\
&& -\alpha \frac{G_N}{1+\alpha} \int d{\bf r}_s \frac{\rho({\bf r}_s)
{\rm e}^{- |{\bf r}-{\bf r}_s|/\lambda}}
{| {\bf r}-{\bf r}_s|} + \mbox{B.C.} \label{pares_eq_gralPsiN}
\end{eqnarray}
The first term of Eq. (\ref{pares_eq_gralPsiN}), is the
contribution of the usual Newtonian gravitation (without SF), 
while information about the SF is contained in the
second term, that is, arising from the influence function determined by the
modified Helmholtz Green function, where the coupling $\omega$ ($\alpha$) enters
as part of a source factor.

\subsection{Cosmological evolution equations using a static STT}
To simulate cosmological systems,  the expansion of the universe has to be
taken into account.
Also, to determine the nature of the cosmological model we need to determine
 the composition of the
universe, i. e., we need to give the values of $\Omega_i$ for each component $i$, 
taking into account
in this way all forms of energy densities that exist at present.
If a particular kind of energy density is described by an equation of state of the form
$p=w \rho$, where $p$ is the pressure and $w$ is a constant, then the equation for energy
conservation in an expanding background, $d(\rho a^3)=-pd(a^3)$, can be integrated to
give $\rho \propto a^{-3(1+w)}$. Then, the Friedmann equation for the expansion factor $a(t)$
is written as
\begin{equation}
\frac{\dot{a}^2}{a^2} = H_0^2 \sum_i \Omega_i \left(\frac{a_0}{a}\right)^{3(1+w_i)} - \frac{k}{a^2}
\end{equation}
where $w_i$ characterizes equation of state of specie $i$. The most familiar forms of energy
densities are those due to pressureless matter with $w_i=0$ (that is, nonrelativistic matter
with rest-mass-energy density $\rho c^2$ dominating over the kinetic-energy density
$\rho v^2/2$) and radiation with $w_i=1/3$.  The density parameter contributed today
by visible, nonrelativistic, baryonic matter in the universe is $\Omega_B \approx (0.01-0.2)$
and the density parameter that is due to radiation is $\Omega_R \approx 2\times 10^{-5}$.
In this work we will consider a model with only two energy density contribution.
One which is a pressureless and
nonbaryonic dark matter  with $\Omega_{DM} \approx 0.3$ that does not couple with radiation.
Other, that will be a cosmological constant contribution $\Omega_\Lambda \approx 0.7$
with and equation of state $p =-\rho$. The above equation for $a(t)$ becomes
\begin{equation}
\frac{\dot{a}^2}{a^2} = H_0^2 
\left[ 
\Omega_{DM} \left(\frac{a_0}{a}\right)^{3} +  \Omega_\Lambda
\right]
- \frac{k}{a^2}
\end{equation}

Here, we employ a cosmological model with a static SF which is consistent with the 
Newtonian limit given by Eq. (\ref{pares_eq_gralPsiN}).
Thus, the scale factor, $a(t)$,  is given by the following Friedman model,
\begin{equation} \label{new_friedman}
a^3 H^2= H_{0}^{2} \left[\frac{\Omega_{m0} +  \Omega_{\Lambda 0} \, a^3}{1+\alpha} 
+  \left(1-\frac{\Omega_{m 0}+\Omega_{\Lambda 0}}{1+\alpha} \right) \, a  \right]
\end{equation}
where $H=\dot{a}/a$,  $\Omega_{m 0}$ and $\Omega_{\Lambda 0}$ 
are the matter and energy density evaluated at present, respectively.   
We notice that the source of the cosmic evolution is deviated by the term 
$1+\alpha$ when compared to the standard Friedman-Lemaitre 
model. Therefore, it is convenient to define a new density parameter by 
$\Omega_i^{(\alpha)} \equiv \Omega_i/(1+\alpha)$. This new density parameter is such that 
$\Omega_m^{(\alpha)} + \Omega_\Lambda^{(\alpha)} =1$, 
which implies a flat universe, and this shall be assumed 
in our following computations, where we consider 
$(\Omega_m^{(\alpha)}, \Omega_\Lambda^{(\alpha)}) = (0.3, 0.7) $.  For positive values 
of $\alpha$, a flat cosmological model demands to have a factor $(1+\alpha)$ more energy 
content ($\Omega_m$ and $ \Omega_\Lambda$) than in standard cosmology. 
On the other hand, for negative values of  
$\alpha$ one needs a factor $(1+\alpha)$  less $\Omega_m$ 
and $ \Omega_\Lambda$ to have a flat universe.  To be consistent 
with the CMB  spectrum and structure formation numerical 
experiments, cosmological constraints must be applied on $\alpha$ in order for it to 
be within the range $(-1,1)$ \cite{Nagata2002,Nagata2003,Shirata2005,Umezu2005}.  

\subsection{The $N$-Body problem for dark matter}
The Vlasov-Poisson equation in an expanding universe describes the evolution of the
six-dimensional, one-particle distribution function, $f(\mathbf{x},\mathbf{p})$.
The Vlasov equation is,
\begin{equation}\label{Vlasov_eq}
\frac{\partial f}{\partial t} + \frac{\mathbf{p}}{m a^2} \cdot \frac{\partial f}{\partial \mathbf{x}} 
-  m \nabla \Phi_N (\mathbf{x}) \cdot \frac{\partial f}{\partial \mathbf{p}} = 0 
\end{equation}
where $\mathbf{x}$ is the comoving coordinate, 
$\mathbf{p}=m a^2 \dot{\mathbf{x}}$, $m$ is the particle mass, and $\Phi_N$ is
the self-consistent gravitational potential given by the
Poisson equation,
\begin{equation}\label{Poisson_eq}
\nabla^2 \Phi_N(\mathbf{x}) = 4 \pi G_N \, a^2 
\left[
\rho(\mathbf{x}) - \rho_b(t)]
\right]
\end{equation}
where $\rho_b$ is the background mass density. 
Eqs. (\ref{Vlasov_eq}) and (\ref{Poisson_eq}) form the Vlasov-Poisson equation,
constitutes a collisionless, mean-field approximation to the evolution of the full
$N$-body distribution. An $N$-body code attempts to solve 
Eqs. (\ref{Vlasov_eq}) and (\ref{Poisson_eq}) by representing the one-particle
distribution function as
\begin{equation}\label{discrete_distribution_eq}
f(\mathbf{x},\mathbf{p}) = \sum_{i=1}^N \delta(\mathbf{x}-\mathbf{x}_i)\, 
\delta(\mathbf{p}-\mathbf{p}_i)
\end{equation}
Substitution of (\ref{discrete_distribution_eq}) in the Vlasov-Poisson system of 
equations yields the exact Newton's equations for a system of $N$ gravitating
particles. See Ref.\ \cite{Bertschinger1998} for details.

In the Newtonian limit of STT of gravity, 
the Newtonian motion equation  for a particle $i$ is written as\cite{mar2008}
\begin{equation} \label{eq_motion}
\ddot{\mathbf{x}}_i + 2\, H \, \mathbf{x}_i = 
-\frac{1}{a^3} \frac{G_N}{1+\alpha} \sum_{j\ne i} \frac{m_j (\mathbf{x}_i-\mathbf{x}_j)}
{|\mathbf{x}_i-\mathbf{x}_j|^3} \; F_{SF}(|\mathbf{x}_i-\mathbf{x}_j|,\alpha,\lambda)
\end{equation}
where the sum includes all  
periodic images of particle $j$,  and $F_{SF}(r,\alpha,\lambda)$ is
\begin{equation}
F_{SF}(r,\alpha,\lambda) = 1+\alpha \, \left( 1+\frac{r}{\lambda} \right)\, e^{-r/\lambda}
\end{equation}
which,  for small distances compared to $\lambda$,  is 
$F_{SF}(r<\lambda,\alpha,\lambda) \approx 1+\alpha \, \left( 1+\frac{r}{\lambda} \right)$ and, 
for long 
distances, is  $F_{SF}(r>\lambda,\alpha,\lambda) \approx 1$, as in Newtonian physics.

\section{Results}
In this section, we present results of cosmological simulations of a $\Lambda$CDM universe
with and without SF contribution. 
We use $256^3$ particles in
box $256 \, h^{-1}$ Mpc size. 
We have studied in the past a $\Lambda$CDM model in a smaller box and with less resolution
than the present case\cite{mar2008}, the $\Lambda$CDM case that comes with 
Gadget 1.0\cite{mar2007}, and the Santa Barbara cluster\cite{mar2009}.

The initial linear power spectrum was generated using the fitting formula by Klypin \& 
Holtzman\cite{KlypinHoltzman1997} for the transfer function. This formula is a slight variation
of the common BBKS fit\cite{Bardeen1986}. It includes effects from baryon suppression but no 
baryonic oscillations. We use the standard Zel'dovich approximation\cite{Zeldovich1970} 
to provide the initial $256^3$ particles displacement off a uniform grid and to assign their initial
velocities
in a $256 \, h^{-1}$ Mpc box.
The starting redshift is $z_{in}=50$ and we choose the following cosmology:
$\Omega_{DM}=0.314$ (where $\Omega_{DM}$ includes cold dark matter and baryons),
$\Omega_{B}=0.044$, $\Omega_{\Lambda}=0.686$, $H_0=71$ km/s/Mpc, $\sigma_8=0.84$,
and $n=0.99$. 
Particle masses are in the order of $1.0\times 10^{10}$ M$_\odot$. 
The individual softening length
was 20 kpc$/h$. This choice of softening length is consistent
with the mass resolution set by the number of particles.
All these values are in concordance with measurements of cosmological
parameters by WMAP\cite{Spergel2003}. 
The initial condition is in the Cosmic Data Bank web page 
(\url{http://t8web.lanl.gov/people/heitmann/test3.html}). 
See Heitmann et al. 2005\cite{Heitmann2005} for more details.

We now present the results for the $\Lambda$CDM model previously described. 
Because the visible component is the smaller one and given our interest to
test the consequences of including a SF contribution to the evolution equations,
our model excludes gas particles, but all its mass has been added to the dark matter. 
We restrict the values of $\alpha$ to the interval $(-1,1)$ 
  \cite{Nagata2002,Nagata2003,Shirata2005,Umezu2005}  and  use $\lambda=1, 5, 10, 20$ 
  Mpc$/h$, since 
these values sweep the scale lengths present in the simulations.
\begin{figure}
\begin{minipage}{6.5in}
\begin{center}
\includegraphics[width=2.in]{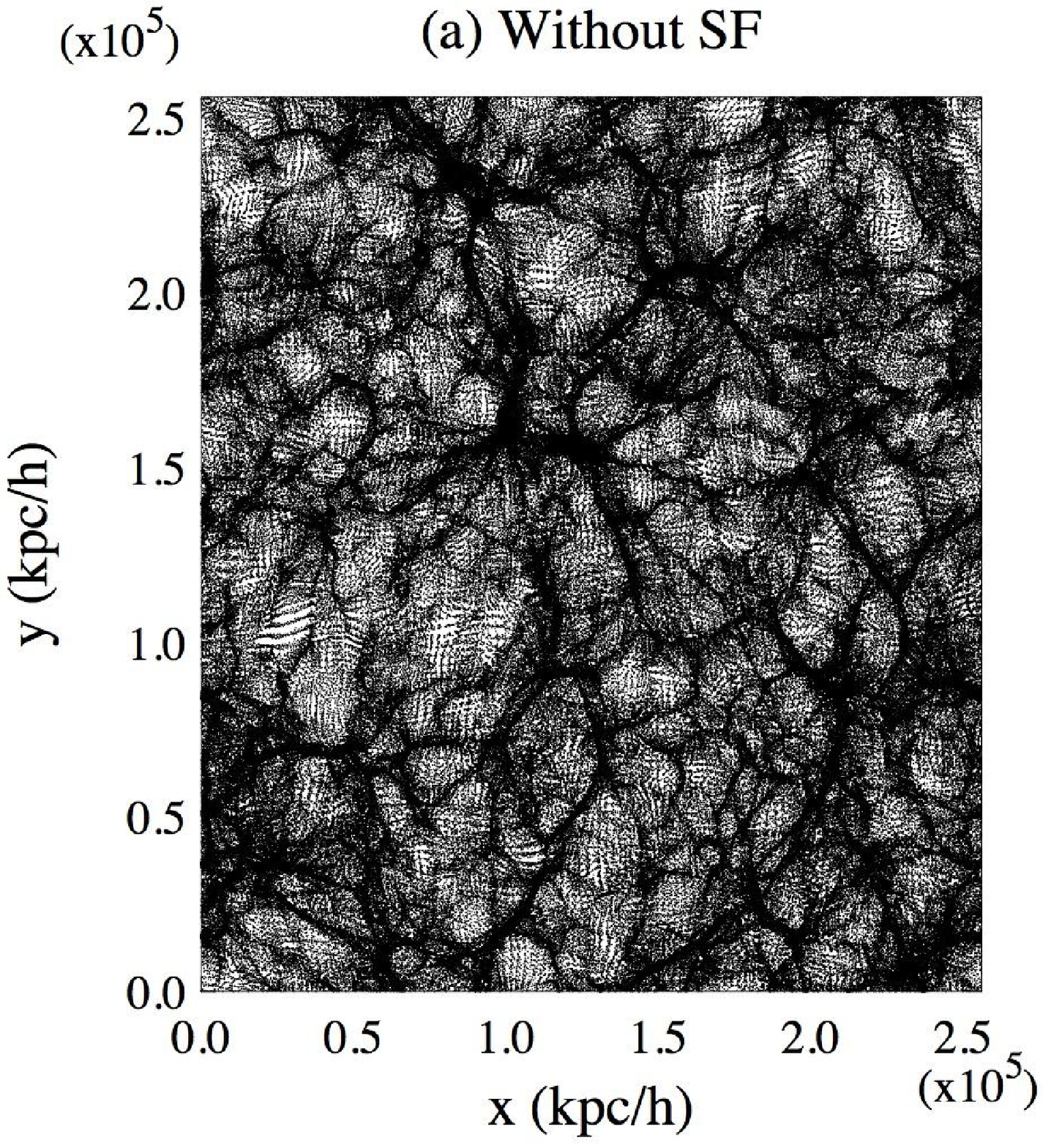} \\
\includegraphics[width=2.in]{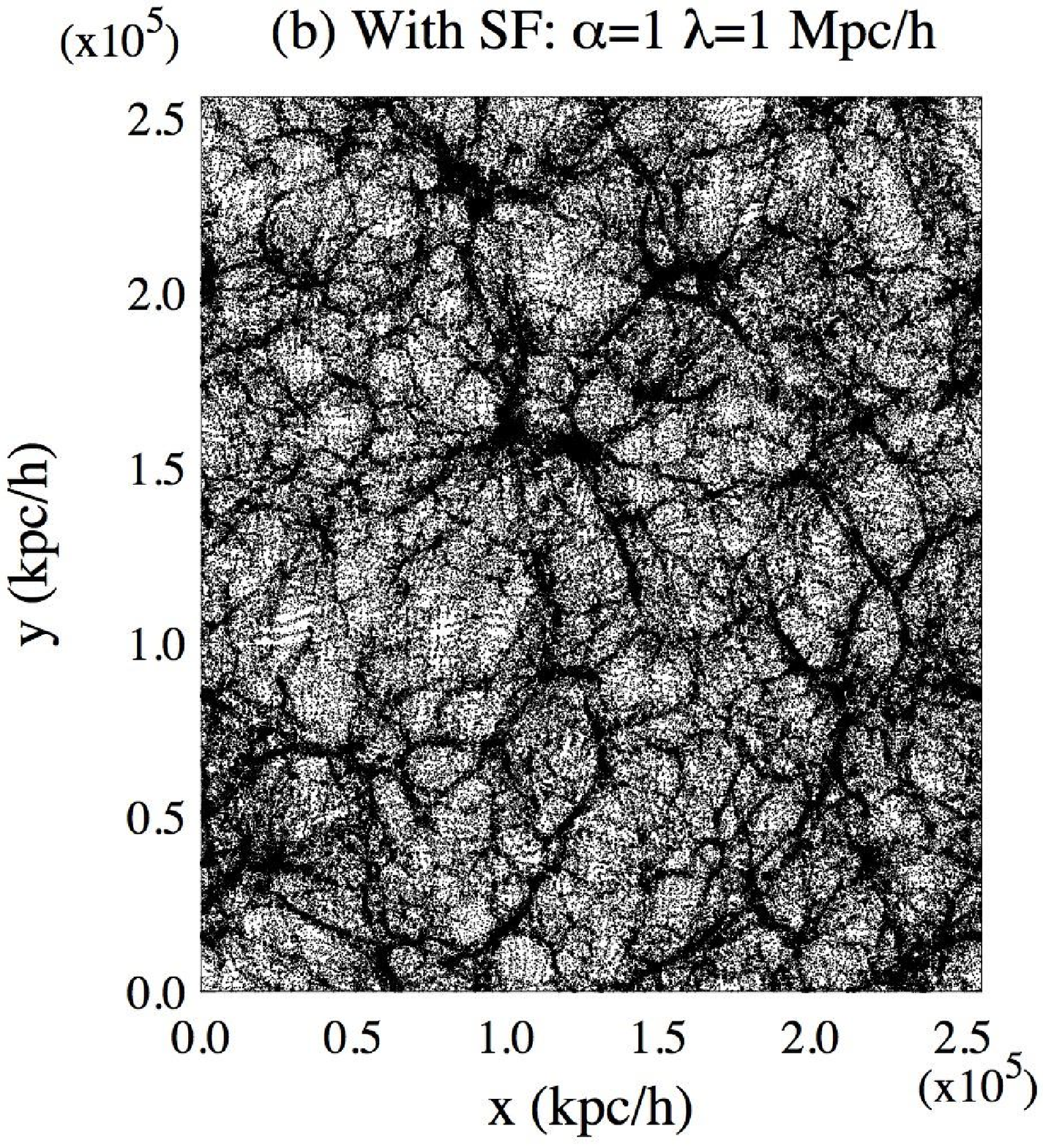}
\includegraphics[width=2.in]{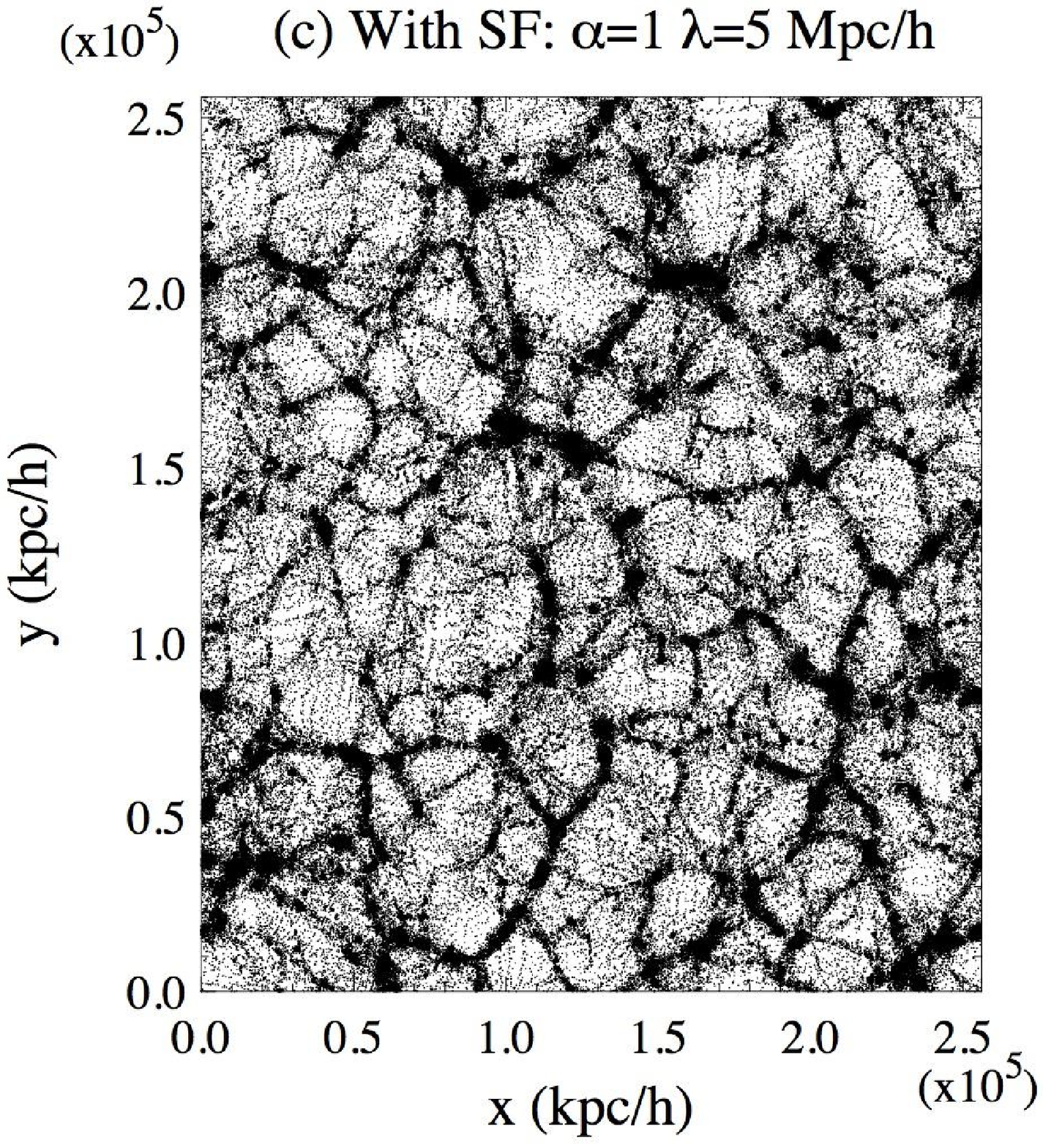}
\includegraphics[width=2.in]{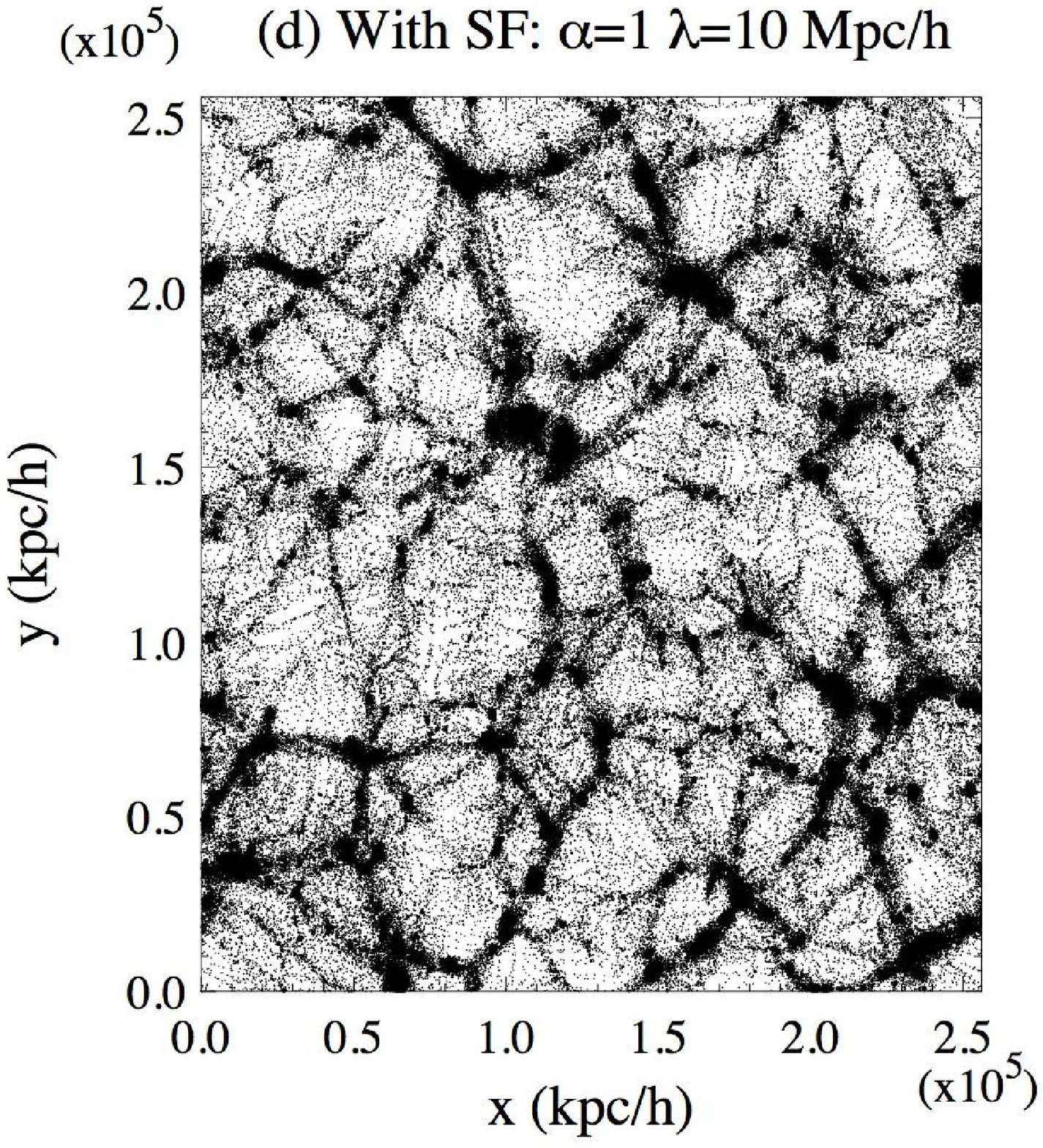}
\includegraphics[width=2.in]{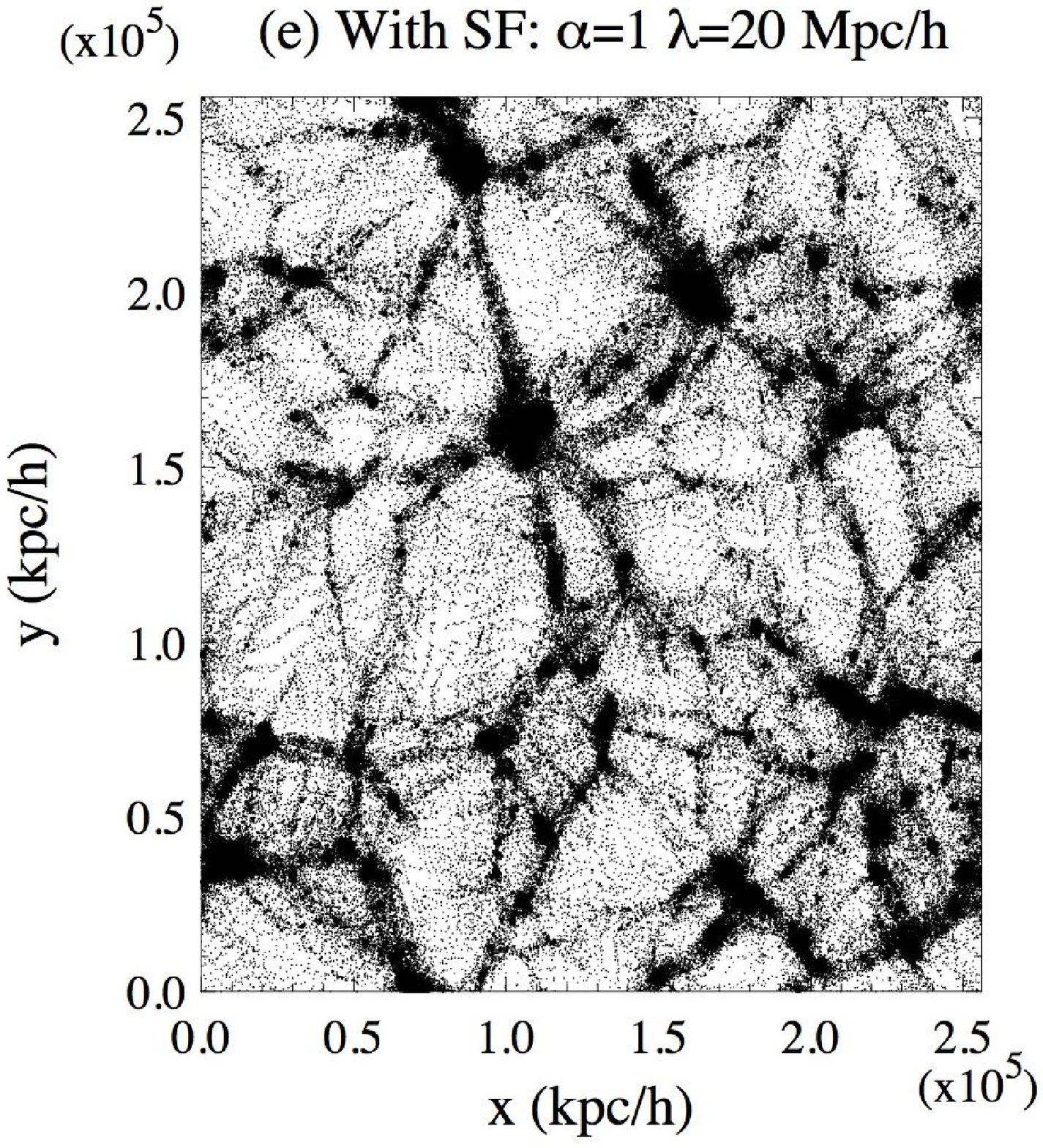}
\includegraphics[width=2.in]{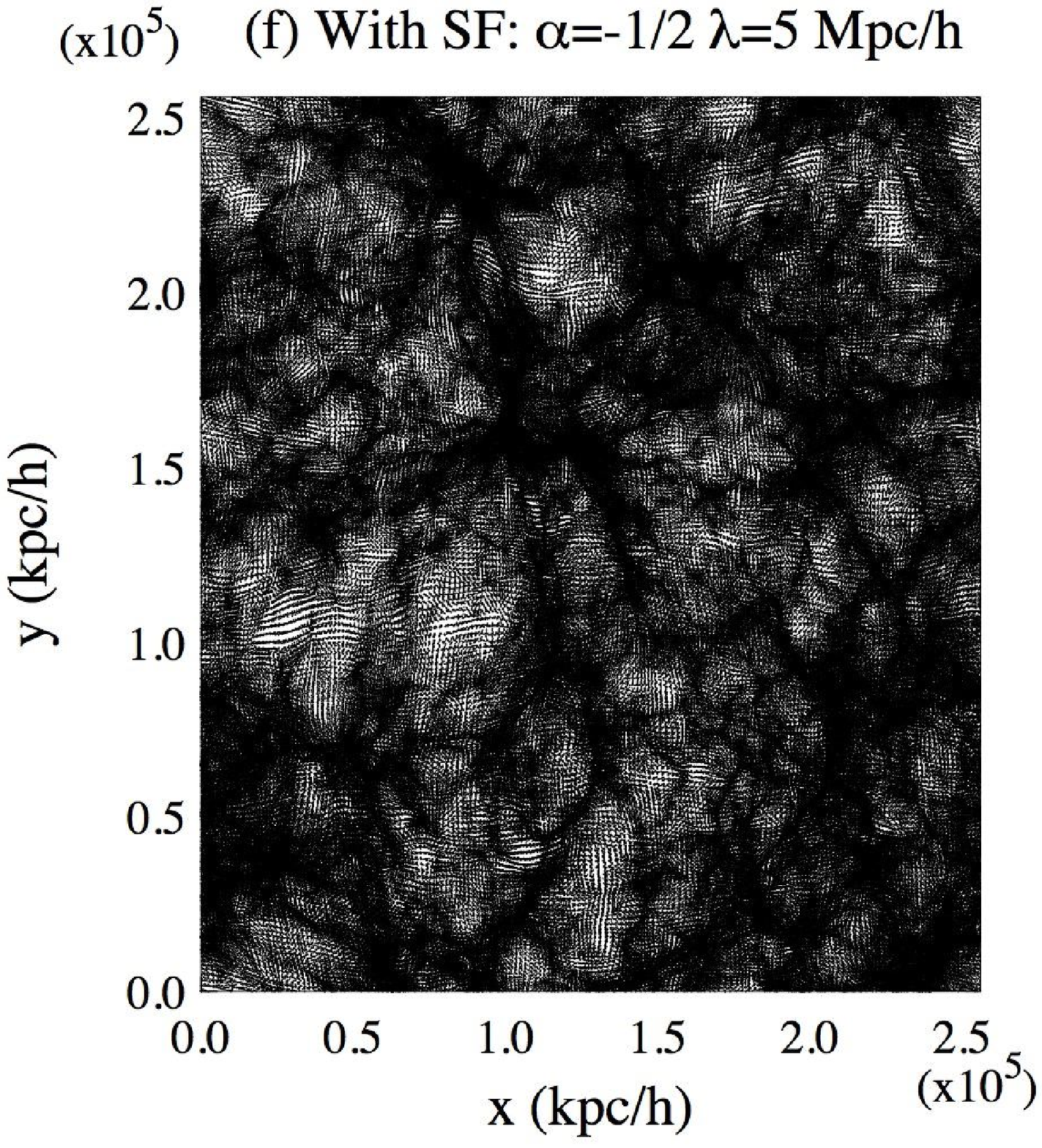}
\includegraphics[width=2.in]{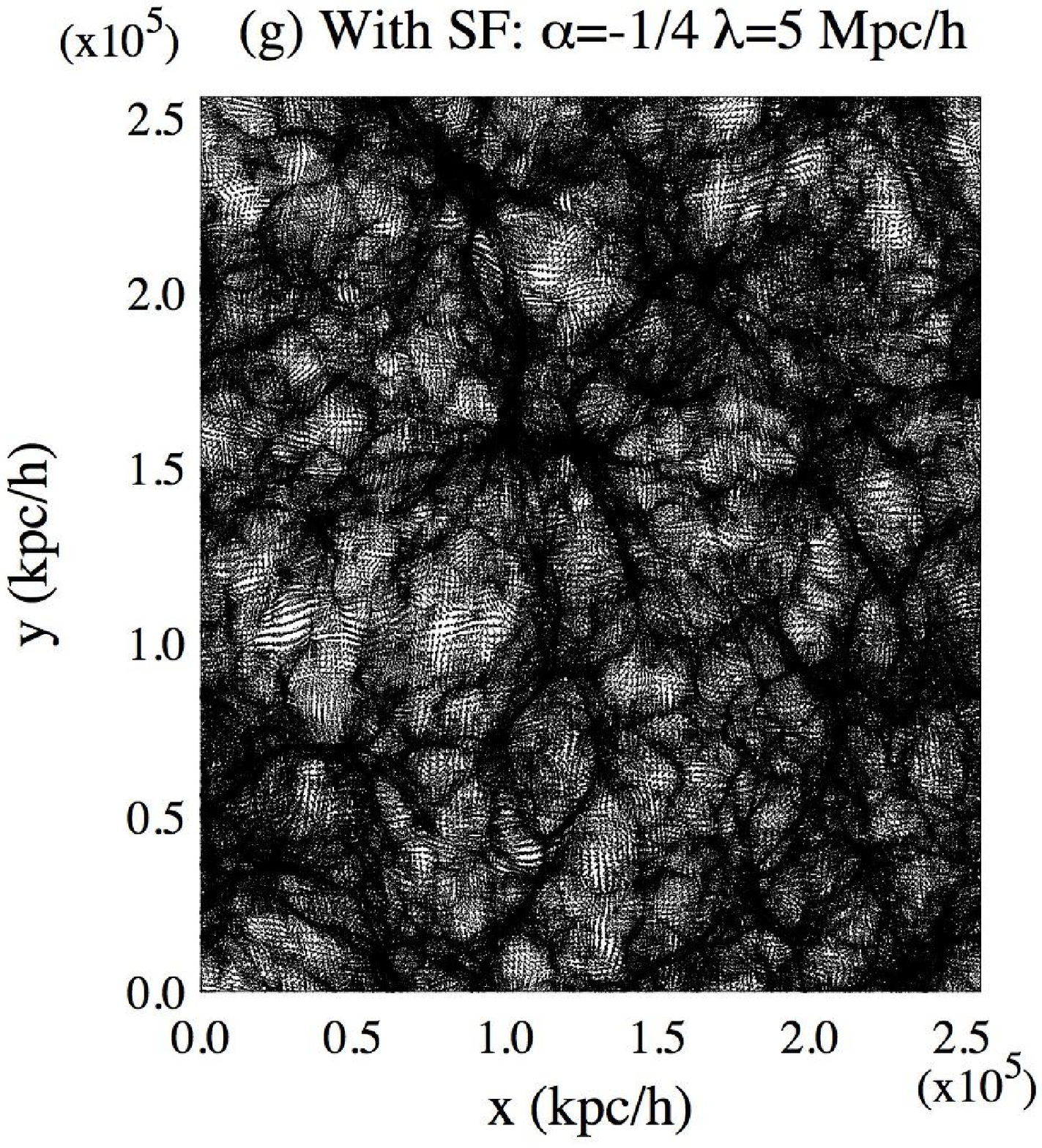}
\end{center}
\end{minipage}
\caption{$x$--$y$ snapshots at $z=0$ of a $\Lambda$CDM universe without and with SF.  
See text for details. }
\end{figure}
In Fig. 1 we show $x$--$y$ snapshots at redshift $z=0$ of our  $\Lambda$CDM model. 
Fig. 1 (a) presents the standard case without SF, i.e., the interaction between bodies  is through
the standard Newtonian potential.
In (b), (c), (d), and (e) we show the case with $\alpha=1$, 
and $\lambda=1, 5, 10, 20$ Mpc$/h$, respectively.
In (f) and (g) $\lambda=5$ Mpc$/h$ and $\alpha=-1/2$ and $-1/4$, respectively.  
One notes clearly how the SF modifies the matter  structure of the system. The most
dramatic cases are (e) and (f) where we have used 
$\alpha=1$ and $\lambda=20$ Mpc$/h$, and $\alpha=-1/2$ and $\lambda=5$ Mpc$/h$, 
respectively. 

We now analyze the general effect that the constant $\alpha$ has on the dynamics.  
The role of $\alpha$ in our approach is
as follows.   On one hand, to construct a flat model  
we have set the condition $\Omega_m^{(\alpha)} + \Omega_\Lambda^{(\alpha)} =1$, which 
implies  having $(1+\alpha)$ times the energy content of the standard $\Lambda$CDM model.
 This essentially means that
we have an increment by a factor of  $(1+\alpha)$ times the amount of matter, 
for positive values of  $\alpha$, or a 
reduction of the same factor for negative values of  $\alpha$. Increasing or reducing 
this amount of matter affects 
the matter term on the  r.h.s. of the equation of 
motion (\ref{eq_motion}), but the amount affected cancels out with the term $(1+\alpha)$ in the 
denominator of  (\ref{eq_motion}) stemming from the new Newtonian potential. 
On the other hand, the factor $F_{SF}$ augments (diminishes) for positive (negative)  
values of $\alpha$ for small distances compared to  $\lambda$, resulting in more (less) structure formation for positive (negative) values of $\alpha$ compared to the $\Lambda$CDM model.  
For $r\gg \lambda$ the dynamics is essentially Newtonian.

Therefore, for cases in which we use $\lambda=5$ Mpc$/h$ we have the following.
In the case of (c), for $r \ll \lambda$,
the effective gravitational pull has been  augmented by a factor of $2$, 
in contrast to case (f) where it has diminished  by a factor of 1/2; in model (g) the pull 
diminishes only by a factor of 3/4. That is why one observes for $r < \lambda$ more structure 
formation in (c), less in (f), and lesser in  model (g).  
The effect is  then, for a growing positive $\alpha$, 
to speed up  the growth of perturbations, then of halos and then of clusters, whereas negative 
$\alpha$ values ($\alpha \rightarrow -1$) tend to slow down the growth. 
Whereas in models (b), (c), (d) and (e) where we keep $\alpha=1$ and take 
$\lambda=1, 5, 10, 20$ Mpc$/h$, we observe less structure in case (b) to more structure in case (e),
passing for intermediate structure formation cases. 
In spite of that the effective gravitational constant has been
augmented by a factor of $2$ the importance of the Yukawa contribution is very small for
distances $r \gg \lambda$. That is way we observe this behavior.

\section{Conclusions}

The theoretical
scheme we have used 
is compatible with local observations because we have defined
the background field constant 
$<\phi>  =  G_{N}^{-1} (1+\alpha)$.  A direct consequence of  
the approach is that  the amount of matter (energy) has to be increased 
for positive values of $\alpha$ and diminished  for negative values of $\alpha$ 
with respect to the standard $\Lambda$CDM model 
in order to have a flat cosmological model. Quantitatively, our model demands to 
have $\Omega/ (1+\alpha) =1$ and this changes the amount of dark matter and 
energy of the model for a flat cosmological model, as assumed.   
The general gravitational effect is that  the interaction including  the SF changes by a factor 
$F_{SF}(r,\alpha,\lambda) \approx 1+\alpha \, \left( 1+\frac{r}{\lambda} \right)$ for $r<\lambda$ in 
comparison with the Newtonian case. Thus, for $\alpha >0$ the growth of structures speeds up  
in comparison with the Newtonian case.  For the   $\alpha <0$ case the effect is to diminish 
the formation of structures.  For $r> \lambda$ the dynamics is essentially Newtonian.
However,  this preliminar analysis we have done is insufficient to give us a clear
conclusions on the role played by SF in the large-scale structure formation process.
We will need to do a systematic study of the evolution of the two-point correlation function which
is a mesure of galaxy clustering.
We also will need to compute the mass power
spectrum and velocity dispersions of the halos. Therefore, we will be able make 
sistematic comparisons with observations. This work is in process and will be published soon.


\bigskip
{\it Acknowledgements: }
This work was supported by CONACYT, grant number I0101/131/07 C-234/07, IAC collaboration. 
The simulations were performed in the UNAM HP cluster {\it Kan-Balam}.

\end{document}